\documentclass[aps,prd,preprintnumbers,nofootinbib,superscriptaddress]{revtex4}
\usepackage{epsfig}
\usepackage{amssymb,amsmath,bbm}
\usepackage{graphicx}
\usepackage{verbatim}
\usepackage{appendix}

\newcommand{\x}{\vec{x}}
\newcommand{\y}{\vec{y}}
\newcommand{\dd}{{\rm d}}

\begin{document}
\preprint{IPMU 15-0057, YITP-15-30}
\title{Hamiltonian analysis of nonprojectable Ho\v{r}ava-Lifshitz gravity with $U(1)$ symmetry}
\author{Shinji Mukohyama}
\affiliation{Yukawa Institute for Theoretical Physics, Kyoto University, Kyoto 606-8502, Japan}
\affiliation{Kavli Institute for the Physics and Mathematics of the Universe (WPI), The University of Tokyo Institutes for Advanced Study, The University of Tokyo, Kashiwa, Chiba 277-8583, Japan}

\author{Ryo Namba}
\affiliation{Kavli Institute for the Physics and Mathematics of the Universe (WPI), The University of Tokyo Institutes for Advanced Study, The University of Tokyo, Kashiwa, Chiba 277-8583, Japan}

\author{Rio Saitou}
\affiliation{Yukawa Institute for Theoretical Physics, Kyoto University, Kyoto 606-8502, Japan}
\affiliation{Institute for Cosmic Ray Research, The University of Tokyo, Kashiwa, Chiba 277-8582, Japan}

\author{Yota Watanabe}
\affiliation{Kavli Institute for the Physics and Mathematics of the Universe (WPI), The University of Tokyo Institutes for Advanced Study, The University of Tokyo, Kashiwa, Chiba 277-8583, Japan}

\begin{abstract}
	We study the nature of constraints and count the number of degrees of
	freedom in the nonprojectable version of the $U(1)$ extension of
	Ho\v{r}ava-Lifshitz gravity, using the standard method of Hamiltonian 
	analysis in the classical field theory. This makes it possible for us to
	investigate the condition under which the scalar graviton is absent at
	a fully nonlinear level. We show that the scalar graviton does not
	exist at the classical level if and only if two specific
	coupling constants are exactly zero. The operators corresponding
	to these two coupling constants are marginal for any values of
	the dynamical critical exponent of the Lifshitz scaling and
	thus should be generated by quantum corrections even if they are
	eliminated from the bare action. We thus conclude that the
	theory in general contains the scalar graviton. 
\end{abstract}

\maketitle

\section{Introduction} 
  \label{sec:introduction}

Ho\v{r}ava~\cite{Horava:2009uw} proposed a power-counting renormalizable
theory of gravity in 2009. What renders the theory renormalizable in the 
power-counting sense is the anisotropic scaling, or Lifshitz scaling, 
\begin{equation}
 t \to b^z t, \quad \vec{x}\to b\vec{x}, \label{eqn:scaling}
\end{equation}
with the dynamical critical exponent $z\geq d$ at high energy, where $d$
is the number of spatial dimensions. Because of this scaling, the theory
is often called Ho\v{r}ava-Lifshitz gravity. The basic quantities 
are the lapse $N$, the shift $N^i$ and the three-dimensional spatial metric
$g_{ij}$. In the minimal theory called the projectable theory, the lapse
is set to be a function of time only. Hence, the basic quantities in the
projectable theory are
\begin{equation}
 N = N(t), \quad N^i = N^i(t,\vec{x}), \quad
  g_{ij} = g_{ij}(t,\vec{x}), \quad (\mbox{projectable}).
\end{equation}
On the other hand, in the nonprojectable theory the lapse may depend
on the spatial coordinates as well as the time, 
and thus the basic quantities yield
\begin{equation}
 N = N(t,\vec{x}), \quad N^i = N^i(t,\vec{x}), \quad
  g_{ij} = g_{ij}(t,\vec{x}), \quad (\mbox{nonprojectable}).
  \label{eqn:nonprojectable}
\end{equation}

The fundamental symmetry of the theory is the invariance under the so-called foliation-preserving diffeomorphism,
\begin{equation}
 t \to t'(t), \quad \vec{x} \to \vec{x}'(t,\vec{x}).
\end{equation}
Since the theory enjoys less symmetries than general relativity (GR), the
number of propagating degrees of freedom is larger. It contains not only
a tensor graviton but also a scalar graviton. The properties of the
scalar graviton at the linearized level crucially depend on a parameter
commonly denoted as $\lambda$. The positivity of the (time) kinetic term 
of the scalar graviton on a flat background requires that
\begin{equation}
 \lambda < \frac{1}{3} \quad \mbox{or} \quad 1 < \lambda.
  \label{eqn:noghost-lambda}
\end{equation}
On the other hand, the dispersion relation of the scalar graviton in the
projectable theory is
\begin{equation}
 \omega^2 = c_s^2 \vec{k}^2 
  + \alpha_2\frac{(\vec{k}^2)^2}{M^2}
  + \alpha_3\frac{(\vec{k}^2)^3}{M^4}, \quad
  c_s^2 = -\frac{\lambda - 1}{3\lambda-1} c_g^2,
  \quad (\mbox{projectable}),
\end{equation}
where $M$ is a mass scale, $\alpha_{2,3}$ are dimensionless
constants and $c_g$ is the speed of gravitational waves. Short-wavelength perturbations are stable, provided that $\alpha_{2,3}$ are
positive. However, for the range of $\lambda$ shown in Eq.~(\ref{eqn:noghost-lambda}), the ``sound speed squared'' $c_s^2$ is
negative and thus the flat background is unstable against long-wavelength perturbations.

One way to tame the instability in the infrared (IR) is to relax the
projectability condition and thus to consider the nonprojectable theory
(\ref{eqn:nonprojectable}). In this case the lapse can depend on
spatial coordinates, and thus terms depending on spatial derivatives of
the lapse may enter the action. This significantly increases the number
of independent coupling constants in the theory but allows for a stable
regime of parameters in the IR. The ``sound speed squared'' is now given
by 
\begin{equation}
 c_s^2 = \frac{\lambda-1}{3\lambda-1}\frac{2c_g^2-\eta}{\eta}c_g^2, 
  \quad (\mbox{nonprojectable}),
\end{equation}
where $\eta$ is the coefficient of the new term 
$g^{ij}\partial_i\ln N\partial_j\ln N$. (The formula for $c_s^2$ in the
projectable theory is recovered in the limit $\eta\to\infty$.) Thus the
scalar graviton is stable in the IR if the condition
(\ref{eqn:noghost-lambda}) is satisfied and if 
\begin{equation}
 0 < \eta < 2c_g^2. 
\end{equation}

Another way out is to keep the projectability condition and to impose
the condition~\cite{Mukohyama:2010xz}
\begin{equation}
 |c_s| \leq \max [\sqrt{|\Phi|}, HL] \ \mbox{for}\ 
  L > \max [0.01\,\mathrm{mm}, 1/M], \quad
  (\mbox{projectable}), \label{eqn:IRcondition}
\end{equation}
where $H$ is the Hubble expansion rate of the background universe, $L$
is the length scale of interest and $\Phi\sim -G_N\rho L^2$ is the 
Newtonian potential. This and the condition (\ref{eqn:noghost-lambda})
imply that $\lambda$ must approach $1$ from above under the
renormalization group (RG) flow as the Hubble expansion rate
decreases. This would pose a nontrivial phenomenological constraint on
properties of the RG flow.

In the projectable theory, in the limit $\lambda\to 1$ deep inside the
condition (\ref{eqn:IRcondition}), the solution to the momentum
constraint equation for scalar-type perturbations around a flat
background becomes singular. Due to this behavior, the perturbative
expansion of the action for the scalar graviton breaks down as $\lambda$
approaches $1$ from above: there is an infinite series of terms in which
the coefficients of terms of higher order in perturbative expansion have
more negative powers of $(\lambda-1)$ and thus are more singular in the
limit. However, all those singular terms are kinetic terms, i.e. terms
with two time derivatives, and terms without time derivatives are always
independent of $\lambda$. For this reason, if the sum of  kinetic terms
can in principle be canonically normalized, then the potential terms
written in terms of the canonically normalized field have only positive
powers of $(\lambda-1)$ and should be regular in the $\lambda\to 1$
limit. In this sense, the canonically normalized scalar graviton is
weakly coupled and expected to be decoupled from the rest of the world
in the $\lambda\to 1$ limit. Indeed, it has been shown for simple cases
such as static, spherically symmetric
configurations~\cite{Mukohyama:2010xz} and superhorizon perturbations
in the expanding universe~\cite{Izumi:2011eh,Gumrukcuoglu:2011ef} that the GR (plus dark matter~\cite{Mukohyama:2009mz}) is
smoothly recovered in the limit if (and only if) nonlinearity is fully
taken into account. For this reason the behavior of the scalar graviton
in the  $\lambda\to 1$ limit does not necessarily pose a conceptual
problem at least in principle. Nonetheless, it is fair to say that the
breakdown of perturbative expansion leads to technical difficulties as
fully nonlinear behavior is not easy to analyze in general, beyond the
above-mentioned explicit examples. As pointed out in Ref.
\cite{Mukohyama:2010xz}, the situation is similar to the Vainshtein
screening mechanism in massive gravity and other modified gravity
theories~\cite{Vainshtein:1972sx,Babichev:2013usa}.

Even in the nonprojectable theory, if we impose observational
constraints on $\lambda$, $\eta$ and $c_g^2$ then one of the following
two should happen: the perturbative expansion breaks down at a scale
considerably lower than the Planck scale~\cite{Papazoglou:2009fj}, or
higher derivative corrections to the dispersion relation must kick
in at an even lower scale~\cite{Blas:2009ck}. The latter case may be in
conflict with observation~\cite{Liberati:2012jf}. Therefore, one might
have to consider the former case in which the perturbative expansion
breaks down at a certain scale. As in the case of the projectable
theory, since the perturbative expansion breaks down only in the kinetic
terms of the scalar graviton, this is not necessarily a problem at least
in principle.
This situation is again similar to the Vainshtein
screening mechanism.

As the third possibility to get around the issue of the scalar graviton,
Ho\v{r}ava and Melby-Thompson~\cite{Horava:2010zj} proposed a $U(1)$
extension of the projectable version of Ho\v{r}ava-Lifshitz gravity in
which the scalar graviton is claimed to be absent. In the context of the
$U(1)$ extension of Ho\v{r}ava-Lifshitz gravity it was originally
claimed that the symmetry of the theory automatically fixes the
parameter $\lambda$ to $1$ and that the scalar graviton is absent
because of this value of $\lambda$. Later it was however shown that the
theory with any value of $\lambda$ can be constructed without spoiling
the symmetry~\cite{daSilva:2010bm}. Nonetheless, at the level of linear
perturbations it was shown that the scalar graviton is still absent for
any $\lambda$. The proof of the absence of the scalar graviton was later
extended to a fully nonlinear level~\cite{Kluson:2010zn}.

While the original theory with the $U(1)$ symmetry is projectable, it is
possible to extend it to a nonprojectable
theory~\cite{Zhu:2011xe,Zhu:2011yu}. At linear perturbations around the
Minkowski background it is known that the  nonprojectable $U(1)$
extension has a scalar graviton for a generic choice of coupling
constants~\cite{Lin:2013tua}. On the other hand, if one of the
parameters in the IR action (which is denoted as $\eta_2$ in the present
paper) is set to zero, then linear perturbations in the theory do not
contain the scalar graviton~\cite{Zhu:2011xe}.

The purpose of the present paper is to perform the Hamiltonian
analysis and to count the number of degrees of freedom in the
nonprojectable version of the $U(1)$ extension. This makes it possible
for us to investigate the condition under which the scalar graviton
disappears at a fully nonlinear level. We shall show that the theory
generically contains the scalar graviton: the scalar graviton does not
exist if and only if two coupling constants are exactly zero. The terms
corresponding to these two specific coupling constants are marginal for
any values of 
the critical exponent $z$ and the spatial dimension $d$,
and thus should be generated by quantum corrections even if they are eliminated from the bare action by hand.

The rest of the present paper is organized as follows. In Sec.
\ref{sec:theory} we review the construction of the $U(1)$ extension of
Ho\v{r}ava-Lifshitz gravity. In Sec.~\ref{sec:hamiltonian
analysis} we adopt the method of Hamiltonian analysis in the classical field
theory to study the nature of constraints in the theory. We then count
the number of degrees of freedom in Sec.~\ref{sec:number of degrees
of freedom}. Section \ref{sec:discussion} is devoted to a summary of
this paper and to some discussions. As supporting materials, we show a
useful technique to ease calculations by using the symmetry of spatial
diffeomorphism in Appendix~\ref{sec:useful relations}, and briefly
discuss the case of the projectable version of the theory in
Appendix~\ref{sec:projectable}.

\section{nonprojectable Ho\v{r}ava-Lifshitz gravity with $U(1)$ symmetry}
\label{sec:theory}

In this section we provide a brief review of the construction of the $U(1)$
extension of Ho\v{r}ava-Lifshitz gravity, following the derivation
presented in Appendix A of Ref. \cite{Lin:2013tua}. For details of the theory
we refer readers to, e.g. Refs. \cite{Zhu:2011xe,Zhu:2011yu}.

The basic quantities of the theory are the lapse $N(t,\x)$, the shift
$N^i(t,\x)$ and the spatial metric $g_{ij}(t, \x)$, supplemented by the
gauge field $A(t,\x)$ and an auxiliary scalar $\nu(t,\x)$, often called the Newtonian prepotential. The transformations of these quantities under
the infinitesimal foliation-preserving diffeomorphism, 
\begin{equation}
 \delta t=f(t), \quad\delta x^i=\xi^i(t,\x),
\end{equation}
are defined as
\begin{align}
  &\delta g_{ij} = f\partial_t g_{ij}+{\cal L}_\xi g_{ij},\quad 
  \delta N^i = \partial_t(N^if)+\partial_t\xi^i+{\cal L}_\xi N^i, \quad
 \delta N = \partial_t(Nf) + \xi^i\partial_iN,
  \nonumber\\ &
  \delta A = \partial_t(Af)+\xi^i\partial_i A, \quad \delta \nu = f\partial_t\nu+\xi^i\partial_i\nu,
  \label{eq:diffeo}
\end{align}
where $\mathcal{L}_\xi$ is the Lie derivative along $\xi^i$. 
On top of the symmetry under the transformation \eqref{eq:diffeo}, we
further introduce an Abelian symmetry, which is thus called the $U(1)$ symmetry, such
that the action of the theory is invariant under the infinitesimal local
transformation parametrized by the gauge parameter $\alpha(t,\x)$. The
basic quantities transform as,
\begin{align}
 \delta N_i = N\partial_i\alpha, \quad \delta A = N\partial_\perp\alpha, \quad \delta \nu = \alpha, \quad \delta g_{ij} = 0, \quad \delta N = 0,
\label{eq:U(1) transform}
\end{align}
where $\partial_\perp\equiv(1/N)(\partial_t-N^i\partial_i)$. Spatial
indices are raised and lowered by $g^{ij}$ and $g_{ij}$,
respectively. The scaling dimensions are then assigned to coordinates
and variables as follows:
\begin{align}
&
 [\partial_i] = 1, \quad[\partial_t] = z, \quad [\dd t \dd^d\vec{x}] = -z - d,
  \quad [\partial_{\perp}] = z, 
\nonumber\\ &
 [g_{ij}] = 0, \quad [N_i] = [N^i] = z-1, \quad [N] = 0,
\nonumber\\ &
 [\alpha] = z-2, \quad [A] = 2z-2, \quad [\nu] = z-2, 
\end{align}
where $d$ is the number of spatial dimensions, and $z \geq d$ is the
dynamical critical exponent, which determines the scaling
(\ref{eqn:scaling}) in the UV regime.

In order to construct a $U(1)$-invariant action, it is convenient to
define the following $U(1)$-invariant ingredients: 
\begin{align}
 \tilde{K}_{ij} &= \frac{1}{2N}(\partial_tg_{ij}-D_i\tilde{N}_j-D_j\tilde{N}_i),
 \nonumber\\
 \tilde{N}^i &= N^i-ND^i\nu,
 \nonumber\\
 \sigma &= \frac{A}{N}-\partial_\perp\nu-\frac{1}{2}D^i\nu D_i\nu ,
\end{align}
where $D_i$ is the spatial covariant derivative compatible with $g_{ij}$.
The scaling dimensions of $\tilde{K}_{ij}$ and $\sigma$ are
\begin{equation}
 [\tilde K_{ij}] = z, \quad [\sigma] = 2z-2. 
\end{equation}
The kinetic action for $g_{ij}$ is then constructed from
$\tilde{K}_{ij}$ as 
\begin{equation}
 \frac{M_{\rm Pl}^2}{2}\int \mathrm{d}t \mathrm{d}^dx \sqrt{g}N
  \left[ \tilde{K}^{ij}\tilde{K}_{ij} - \lambda\tilde{K}^2 \right], 
\end{equation}
where $M_{\rm Pl}$ is a constant corresponding to the Planck scale at low energy, $\lambda$ is a dimensionless constant,
$g\equiv\mathrm{det}(g_{ij})$ and $\tilde{K}\equiv g^{ij}\tilde{K}_{ij}$. The
scaling dimension of the kinetic terms is 
\begin{equation}
 [\tilde{K}^{ij}\tilde{K}_{ij}] = [\tilde{K}^2] = 2z,
\end{equation}
and thus the power-counting renormalizability requires that the other terms
in the action should have scaling dimensions equal to or less than
$2z$. In particular, it is required that the dependence of the action on
$\sigma$ be at most linear. This is because 
\begin{equation}
 [\sigma^n] = 2n(z-1) > 2z, \quad 
  \mbox{for} \ n\geq 2, \ z\geq d \geq 3 ,
\end{equation}
and because derivatives and all other fundamental variables have
non-negative scaling dimensions in $z \geq d \geq 3$. On the other hand,
the linear term, $\sigma$, is always power-counting renormalizable
(relevant) since $[\sigma]=2z-2<2z$. Other terms linear in $\sigma$,
that is $R\sigma$, $a^i a_i \sigma$, and $D^i a_i \sigma$, are marginal,
i.e., 
\begin{equation}
 [R\sigma] = [a^ia_i\sigma] = [D^ia_i\sigma] = 2z,
\end{equation}
where $R$ is the Ricci scalar of $g_{ij}$ and 
$a_i\equiv D_i\ln N$. 
We do not include in the action the time derivatives of $N$, $N^i$ or
$A$, since no relevant/marginal terms that are invariant under both the
foliation-preserving diffeomorphism and the $U(1)$ transformation can be
constructed.

The gravity action which respects foliation-preserving diffeomorphism,
the $U(1)$ symmetry and the power-counting renormalizability in UV thus reads 
\begin{equation}
 I = \frac{M_{\rm Pl}^2}{2}\int \mathrm{d}t \mathrm{d}^dx \sqrt{g}N
  \left[ \tilde{K}^{ij}\tilde{K}_{ij} - \lambda\tilde{K}^2 -
   \mathcal{L}_V + (2\Omega-\eta_0R+\eta_1a^ia_i+\eta_2D^ia_i)\sigma
  \right], \label{eq:action}
\end{equation}
where $\Omega$ and $\eta_{0,1,2}$ are constants, and the potential 
term $\mathcal{L}_V[g_{ij},N]$ is constructed from $a_i$, $g_{ij}$,
$D_i$ and the Riemann tensor of $g_{ij}$, including all the terms that
contain up to $2z$ spatial derivatives and are invariant under the
diffeomorphism. One of the coupling constants $\eta_{0,1,2}$ is in fact
redundant and one can, for example, set
\begin{equation}
 \eta_0 = 1 ,
\end{equation}
by a redefinition of $A$ and $\Omega$.

\section{Hamiltonian analysis}
\label{sec:hamiltonian analysis}

It has been shown that the projectable version, i.e. the lapse $N$ being 
a function only of time $t$, of the $U(1)$ extension of
Ho\v{r}ava-Lifshitz gravity has the same number of degrees of freedom as
GR, or equivalently one less degree of freedom than the original version
of Ho\v{r}ava-Lifshitz theory
\cite{Horava:2010zj,daSilva:2010bm,Kluson:2010zn}. 
However, once the nonprojectable version, i.e. $N = N(t , \vec x)$, is considered with all the terms consistent with the symmetry included, an additional scalar degree of freedom, dubbed a scalar graviton, reappears already at the level of linearized perturbations \cite{Lin:2013tua}. In this section, in order to understand the complete structure of the theory, we perform the Hamiltonian analysis and count the number of degrees of freedom at the fully nonlinear order.
Since we are interested in the physical degrees of freedom contained in the gravity sector, we shall omit matter fields.

We start with the $(d^2+3d+6)$-dimensional phase space ($g_{ij}$,
$\pi^{ij}$, $N^i$, $\pi_i$, $N$, $\pi_N$, $A$, $\pi_A$, $\nu$,
$\pi_\nu$) at each point in spacetime,
where $\pi^{ij}$, $\pi_i$, $\pi_N$, $\pi_A$
and $\pi_\nu$ are the canonical momenta conjugate to $g_{ij}$, $N^i$,
$N$, $A$ and $\nu$, respectively. The Poisson bracket is defined in the
standard way as 
\begin{align}
 \left\{F,G\right\}_{\rm P} \equiv& \int \mathrm{d}^dx \biggl[ 
 \frac{\delta F}{\delta g_{ij}(\x)}\frac{\delta G}{\delta \pi^{ij}(\x)} 
 + \frac{\delta F}{\delta N^i(\x)}\frac{\delta G}{\delta \pi_i(\x)}
 + \frac{\delta F}{\delta N(\x)}\frac{\delta G}{\delta \pi_N(\x)} 
\nonumber\\
&\qquad\quad
 + \frac{\delta F}{\delta A(\x)}\frac{\delta G}{\delta \pi_A(\x)} 
 + \frac{\delta F}{\delta \nu(\x)}\frac{\delta G}{\delta \pi_\nu(\x)} 
 - \frac{\delta F}{\delta \pi^{ij}(\x)}\frac{\delta G}{\delta g_{ij}(\x)}
 \nonumber\\
 & \qquad\quad
 - \frac{\delta F}{\delta \pi_i(\x)}\frac{\delta G}{\delta N^i(\x)} 
- \frac{\delta F}{\delta \pi_N(\x)}\frac{\delta G}{\delta N(\x)} 
- \frac{\delta F}{\delta \pi_A(\x)}\frac{\delta G}{\delta A(\x)} 
- \frac{\delta F}{\delta \pi_\nu(\x)}\frac{\delta G}{\delta \nu(\x)} \biggr].
\end{align}
The canonical momenta are found as
\begin{align}
 &\pi_i=0, \quad \pi_N=0, \quad \pi_A=0, \quad \pi_\nu=-J_A,
  \label{eq:primary}\\
 &\pi^{ij}=\frac{M_{\rm Pl}^2}{2}\sqrt{g}\left[\frac{1}{2}(g^{ik}g^{jl}+g^{il}g^{jk})-\lambda g^{ij}g^{kl}\right]\tilde{K}_{kl},
  \label{eq:pi^{ij}}
\end{align}
where
\begin{equation}
 J_A \equiv \frac{M_{\rm Pl}^2}{2}\sqrt{g}(2\Omega-\eta_0 R+\eta_1a^ia_i+\eta_2D^ia_i).
  \label{def-JA}
\end{equation}
Notice that Eq.~\eqref{eq:primary} constitutes the primary constraints of the system.
We can invert the relation \eqref{eq:pi^{ij}} to express $\tilde K_{ij}$ in terms of $\pi^{ij}$ as
\begin{equation}
 \tilde{K}_{ij}=\frac{2}{M_{\rm Pl}^2\sqrt{g}}\mathcal{G}_{ijkl}\pi^{kl},
\end{equation}
where for notational brevity we have defined
\begin{equation}
 \mathcal{G}_{ijkl} \equiv \frac{1}{2}(g_{ik}g_{jl}+g_{il}g_{jk})-\frac{\lambda}{d\lambda-1}g_{ij}g_{kl} \; .
\end{equation}
As is clear from Eq.~\eqref{eq:primary}, such an inversion cannot be done for $\partial_tN^i$,
$\partial_tN$, $\partial_tA$ and $\partial_t\nu$. While the action
\eqref{eq:action} does not contain $\partial_t N^i$, $\partial_t N$ or
$\partial_t A$
and so there is no need for the inversion,
$\partial_t \nu$ does not enter the Hamiltonian either
once the constraint $\pi_\nu = - J_A$ in Eq.~\eqref{eq:primary} is
imposed. One can thus see that the $(d+3)$ primary constraints 
$\pi_i \approx \pi_N \approx \pi_A \approx \pi_\nu +J_A \approx 0$ are
all independent.

To proceed, we first inspect the algebra of the primary constraints. 
The Poisson brackets of $\pi_i$ and $\pi_A$ with all the constraints
trivially vanish:
\begin{equation}
 \left\{\pi_i(\x),\Phi(\vec{y})\right\}_{\rm P} = 0, \quad \left\{\pi_A(\x),\Phi(\vec{y})\right\}_{\rm P} = 0,
\end{equation}
where $\Phi$ stands for $\{\pi_i,\pi_A,\pi_N,\pi_\nu+J_A\}$. The Poisson
brackets between $\pi_N$ and $\pi_\nu+J_A$ are, on the other hand, 
\begin{align}
&\left\{\pi_N(\x),\pi_N(\vec{y})\right\}_{\rm P} = \left\{\pi_\nu(\x)+J_A(\x),\pi_\nu(\vec{y})+J_A(\vec{y})\right\}_{\rm P} = 0, \\
&\left\{\overline{\pi_N}[\chi],\overline{
\pi_\nu+J_A}[\varphi]\right\}_{\rm P} = \left\{\overline{\pi_N}[\chi],\overline{J_A}[\varphi]\right\}_{\rm P}
 = -\frac{M_{\rm Pl}^2}{2}\int\mathrm{d}^dx \sqrt{g}\varphi\left(\eta_2D^2+2\eta_1a_iD^i\right)\frac{\chi}{N},
  \label{eq:pi_n and J_A}
\end{align}
where the over-bar denotes
\begin{equation}
 \overline{\Phi}[\chi] \equiv \int \mathrm{d}^dx \, \Phi (\vec x) \chi (\vec x) \; ,
 \label{def-phibar}
\end{equation}
where $\Phi$ is any constraint of scalar type and $\chi$ is an arbitrary function. Hereafter we assume that the arbitrary functions corresponding to $\chi(\vec x)$ are independent of the canonical variables; otherwise, some of the strong equalities ($=$) should be replaced by the weak ones ($\approx$).

The Hamiltonian of the system is now constructed as 
\begin{align}
 H \equiv & \int \mathrm{d}^dx [\pi^{ij}\partial_tg_{ij}+\pi_i\partial_tN^i+\pi_N\partial_tN+\pi_A\partial_tA+\pi_\nu\partial_t\nu-\mathcal{L}]
 \nonumber\\
 =&
\int \dd^d x \bigg[
 N \sqrt{g} \left( \frac{2}{M_{\rm Pl}^2} \, \frac{\pi^{ij}}{\sqrt{g}} {\cal G}_{ijkl} \frac{\pi^{kl}}{\sqrt{g}} 
+ \frac{M_{\rm Pl}^2}{2} {\cal L}_V \right)
- N {\cal H}_i D^i \nu
+ \frac{N}{2} J_A D_i \nu D^i \nu
 \nonumber\\
 & \qquad\quad
 + N^i \left( {\cal H}_i - J_A D_i \nu \right) - A J_A
 \bigg],
  \label{def-H}
\end{align}
where $\mathcal{L}$ is the Lagrangian density and we have defined 
\begin{equation}
 \mathcal{H}_i\equiv-2\sqrt{g}D_j \left(\frac{\pi^j_{\; i}}{\sqrt{g}} \right).
\end{equation}
Because of the presence of the constraints, the time evolution in the
phase space is not completely determined by this Hamiltonian. Including
the terms corresponding to the primary constraints, we thus redefine the
Hamiltonian as
\begin{equation}
 \hat{H} = H + \int \mathrm{d}^dx \left[ \lambda^i\pi_i+\lambda_A\pi_A+\lambda_N\pi_N+\lambda_J(\pi_\nu+J_A) \right],
\end{equation}
where $\lambda^i$, $\lambda_A$, $\lambda_N$, $\lambda_\nu$ are Lagrange
multipliers.

Now we need to impose the consistency conditions for the primary
constraints against time evolution with respect to $\hat H$ to determine
the secondary constraints. However, we observe from Eq.
\eqref{eq:pi_n and J_A} that the constraint structure differs depending
on whether the right-hand side of the equation vanishes identically or
not. We therefore separate the analysis into the following two
subsections, one corresponding to the generic case where the right-hand
side of Eq.~\eqref{eq:pi_n and J_A} is nonvanishing, and the other to the
special case where it is identically zero.

\subsection{Generic case}
\label{subsec:generic}

In this subsection we consider the case where the right-hand side of
Eq.~\eqref{eq:pi_n and J_A} does not vanish identically, namely the case
with ($\eta_1,\eta_2)\ne(0,0)$. We call it generic since even if
$\eta_1$ and $\eta_2$ are zero at the classical level, they generically
become nonzero when quantum corrections to the action are taken into
account.

The consistency conditions of the primary constraints with the time
evolution are
\begin{align}
 0 &= \frac{\mathrm{d}}{\mathrm{d}t}(\pi_\nu+J_A) \approx \left\{\pi_\nu+J_A,\hat{H}\right\}_{\rm P} = \left\{\pi_\nu+J_A,H\right\}_{\rm P}+\left\{\pi_\nu+J_A,\overline{\pi_N}[\lambda_N]\right\}_{\rm P},
 \nonumber\\
 0 &= \frac{\mathrm{d}}{\mathrm{d}t}\pi_N \approx \left\{\pi_N,\hat{H}\right\}_{\rm P} = \left\{\pi_N,H\right\}_{\rm P} + \left\{\pi_N,\overline{\pi_\nu+J_A}[\lambda_J]\right\}_{\rm P},
 \label{eqn:consistency-generic-1}
\end{align}
and
\begin{align}
 0 &= \frac{\mathrm{d}}{\mathrm{d}t}\pi_A \approx \left\{\pi_A,\hat{H}\right\}_{\rm P} = \left\{\pi_A,H\right\}_{\rm P} = -\frac{\delta H}{\delta A} = J_A,
 \nonumber\\
 0 &= \frac{\mathrm{d}}{\mathrm{d}t}\pi_i \approx \left\{\pi_i,\hat{H}\right\}_{\rm P} = \left\{\pi_i,H\right\}_{\rm P} = -\frac{\delta H}{\delta N^i} = -\mathcal{H}_i + J_AD_i\nu,
 \label{eqn:consistency-generic-2}
\end{align}
where $\approx$ denotes an equality in the weak sense, i.e.\ with all
the constraints imposed. The first two equations
(\ref{eqn:consistency-generic-1}) do not lead to additional constraints
but are simply considered as equations to determine the multipliers
$\lambda_N$ and $\lambda_J$, 
\begin{align}
 \eta_2D^2\frac{\lambda_N}{N} + 2\eta_1a_iD^i\frac{\lambda_N}{N} \approx& -\frac{2}{M_{\rm Pl}^2\sqrt{g}}\left\{\pi_\nu+J_A,H\right\}_{\rm P},
 \nonumber\\
 \eta_2D^2\lambda_J - 2\eta_1D^i(a_i\lambda_J) \approx& \frac{2N}{M_{\rm Pl}^2\sqrt{g}}\left\{\pi_N,H\right\}_{\rm P}.
  \label{eq:multipliers}
\end{align}
On the other hand, the last two equations
(\ref{eqn:consistency-generic-2}) give secondary constraints,
$J_A\approx 0$ and $\mathcal{H}_i\approx 0$.

The set of constraints that we obtained so far is ($\pi_i$, $\pi_A$,
$\pi_\nu$, $\mathcal{H}_i$, $\pi_N$, $J_A$). Let us calculate the
Poisson brackets among them. The Poisson brackets of $\pi_i$,  $\pi_A$
and $\pi_\nu$ with all the constraints trivially vanish:  
\begin{equation}
 \left\{\pi_i(\x),\Phi(\vec{y})\right\}_{\rm P} = 0, \quad \left\{\pi_A(\x),\Phi(\vec{y})\right\}_{\rm P} = 0, \quad \left\{\pi_\nu(\x),\Phi(\vec{y})\right\}_{\rm P} = 0,
\end{equation}
where $\Phi=\{\pi_j,\pi_A,\pi_\nu,\mathcal{H}_j,\pi_N,J_A\}$. On the
other hand, the Poisson bracket between $\mathcal{H}_i$ and $J_A$ can be
calculated by using the formula
\begin{equation}
 \left\{\overline{\mathcal{H}}[f],F\right\}_{\rm P}
  = 
  \int \mathrm{d}^dx 
  \left( \frac{\delta F}{\delta s}f^i\partial_is 
   + \frac{\delta F}{\delta V^i}[f,V]^i \right),
  \label{eq:formula}
\end{equation}
where $F=F[g_{ij},\pi^{ij},s(\mbox{scalar}),V^i(\mbox{vector})]$ is a 
functional invariant under time-independent spatial diffeomorphism and 
$[f,V]^i=f^jD_jV^i-V^jD_jf^i$ (see Appendix \ref{sec:useful relations}
for the proof). The result is 
\begin{align}
 \left\{\overline{\mathcal{H}}[f],\overline{J_A}[\varphi]\right\}_{\rm P} =& \overline{J_A}[f\partial\varphi] + \int\mathrm{d}^dx \frac{\delta\overline{J_A}[\varphi]}{\delta N}f^i\partial_i N
 \nonumber\\
 \approx& -\left\{\overline{\pi_N}[f\partial N],\overline{J_A}[\varphi]\right\}_{\rm P},
 \label{H-JA}
\end{align}
and this does not vanish weakly, where
\begin{align}
 \overline{\mathcal{H}}[f] \equiv  \int \mathrm{d}^dx \mathcal{H}_if^i, \quad f\partial\varphi \equiv f^i\partial_i\varphi.
\end{align}
Similarly to the comment made after Eq.~\eqref{def-phibar}, we assume that the functions $f^i(\vec x)$ are independent of the canonical variables; otherwise, the strong equality in the first line of Eq.~\eqref{H-JA} should be replaced by the weak one, and the same applies to the following calculations.
We thus define the linear combination 
\begin{equation}
 \mathcal{H}_i^N \equiv \mathcal{H}_i + \pi_N D_iN,
  \label{eqn:def-HNi}
\end{equation}
so that the Poisson bracket of $\mathcal{H}^N_i$ with $J_A$ vanishes
weakly. Actually, the Poisson brackets of $\mathcal{H}^N_i$ with all the
constraints vanish weakly: 
\begin{align}
 &\left\{\overline{\mathcal{H}^N}[f],\overline{\mathcal{H}^N}[g]\right\}_{\rm P} = \overline{\mathcal{H}^N}[[f,g]] \approx 0 \quad \text{for} \quad^\forall f^i, \ ^\forall g^i,
 \nonumber\\
 &\left\{\overline{\mathcal{H}^N}[f],\overline{\Phi}[\varphi]\right\}_{\rm P} = \overline{\Phi}[f\partial\varphi] \approx 0 \quad \text{for} \quad^\forall f^i, \ ^\forall \varphi,
\end{align}
where $\Phi=\{\pi_N,J_A\}$. The remaining Poisson brackets are those
among $\pi_N$ and $J_A$ and are 
\begin{align}
 &\left\{\pi_N(\x),\pi_N(\vec{y})\right\}_{\rm P} = 0, \quad  \left\{J_A(\x),J_A(\vec{y})\right\}_{\rm P} = 0,
\end{align}
and Eq.~\eqref{eq:pi_n and J_A}.

Hereafter we consider ($\pi_i$, $\pi_A$, $\pi_\nu$, $\mathcal{H}_i^N$,
$\pi_N$, $J_A$) as the basis for the set of constraints. Adding the
terms corresponding to the secondary constraints, we thus redefine the
Hamiltonian as 
\begin{equation}
 \tilde{H} = H + \int \mathrm{d}^dx \left[ \lambda^i\pi_i+\lambda_A\pi_A+\lambda_\nu\pi_\nu+\lambda_\mathcal{H}\mathcal{H}_i^N+\lambda_N\pi_N+\lambda_JJ_A \right],
\end{equation}
where $\lambda^i$, $\lambda_A$, $\lambda_\nu$, $\lambda_\mathcal{H}$,
$\lambda_N$, $\lambda_J$ are Lagrange multipliers.

The consistency conditions of $J_A$, $\pi_N$ with the time evolution
with respect to the Hamiltonian $\tilde{H}$ are
\begin{align}
 0 &= \frac{\mathrm{d}}{\mathrm{d}t}J_A \approx  \left\{J_A,H\right\}_{\rm P}+\left\{J_A,\overline{\pi_N}[\lambda_N]\right\}_{\rm P},
 \nonumber\\
 0 &= \frac{\mathrm{d}}{\mathrm{d}t}\pi_N \approx \left\{\pi_N,H\right\}_{\rm P} + \left\{\pi_N,\overline{J_A}[\lambda_J]\right\}_{\rm P}.
\end{align}
They yield no additional constraint but equations to determine the multipliers $\lambda_N$ and $\lambda_J$ as Eq.~\eqref{eq:multipliers}. The consistency conditions of $\pi_A$, $\pi_i$, $\pi_\nu$ and $\mathcal{H}^N_i$ with the time evolution are
\begin{align}
 0 &= \frac{\mathrm{d}}{\mathrm{d}t}\pi_A \approx \left\{\pi_A,H\right\}_{\rm P} = -\frac{\delta H}{\delta A} = J_A,
 \nonumber\\
 0 &= \frac{\mathrm{d}}{\mathrm{d}t}\pi_i \approx \left\{\pi_i,H\right\}_{\rm P} = -\frac{\delta H}{\delta N^i} = -\mathcal{H}_i+ J_AD_i\nu,
 \nonumber\\
 0 &= \frac{\mathrm{d}}{\mathrm{d}t}\pi_\nu \approx \left\{\pi_\nu,H\right\}_{\rm P} = -\frac{\delta H}{\delta \nu} = -\partial_i(N\mathcal{H}^i+\tilde{N}^iJ_A),
 \nonumber\\
 0 &= \frac{\mathrm{d}}{\mathrm{d}t}\overline{\mathcal{H}^N}[f] \approx \left\{\overline{\mathcal{H}^N}[f],H\right\}_{\rm P}
 \nonumber\\
 &= 
 \int \dd^dx
 \left[
 -(N\mathcal{H}^i+\tilde{N}^iJ_A)\partial_i(f\partial\nu)-J_Af\partial A
 + (\mathcal{H}_i-J_A\partial_i\nu)[f,N]^i \right],
\end{align}
where we have used the formula (\ref{eq:formula}) to derive the last
equality. They vanish weakly and thus do not give any additional
constraints.

In summary, ($\pi_i$, $\pi_A$, $\pi_\nu$, $\mathcal{H}_i^N$, $\pi_N$,
$J_A$) forms the set of all constraints for the generic case studied in
this subsection. Among them, ($2d+2$) constraints ($\pi_i$,
$\mathcal{H}_i^N$, $\pi_A$, $\pi_\nu$) are first class and two
constraints ($\pi_N$, $J_A$) are second class.

\subsection{Special case with $\eta_1=\eta_2=0$}
\label{subsec:special}

In this subsection we consider the special case where the right-hand
side of Eq.~\eqref{eq:pi_n and J_A} vanishes identically, namely the case
with $\eta_1=\eta_2=0$. We consider that this is not generic but
rather exceptional. Indeed, the choice of parameters $\eta_1=\eta_2=0$
is not stable under radiative corrections. We nonetheless investigate this
case for completeness. Also, the study of this special case sheds some
light on important roles of Eq.~\eqref{eq:multipliers} for the structure of
the theory: with (and only with) $\eta_1=\eta_2=0$,
Eq.~\eqref{eq:multipliers} leads to secondary constraints instead of yielding
equations to determine the Lagrange multipliers.

The consistency conditions of the primary constraints with the time evolution are
\begin{align}
 0 &= \frac{\mathrm{d}}{\mathrm{d}t}\pi_A \approx \left\{\pi_A,H\right\}_{\rm P} = -\frac{\delta H}{\delta A} = J_A,
 \nonumber\\
 0 &= \frac{\mathrm{d}}{\mathrm{d}t}\pi_i \approx \left\{\pi_i,H\right\}_{\rm P} = -\frac{\delta H}{\delta N^i} = -\mathcal{H}_i+ J_AD_i\nu,
 \nonumber\\
 0 &= \frac{\mathrm{d}}{\mathrm{d}t}\pi_N \approx  \left\{\pi_N,H\right\}_{\rm P} = -\frac{\delta H}{\delta N} = -\mathcal{H}_\perp + \mathcal{H}^iD_i\nu - \frac{1}{2} J_AD^i\nu D_i\nu,
 \nonumber\\
 0 &= \frac{\mathrm{d}}{\mathrm{d}t}(\pi_\nu+J_A) \approx - \frac{\delta H}{\delta \nu} + \left\{J_A,H\right\}_{\rm P}
 \nonumber\\
 &= -\partial_i(N\mathcal{H}^i+\tilde{N}^iJ_A) + \phi_A +
 \frac{2N}{M_{\rm Pl}^2}g^{ij}\mathcal{G}_{ijkl}\pi^{kl}\frac{J_A}{\sqrt{g}} 
+ \sqrt{g}D_i\left(\tilde{N}^i\frac{J_A}{\sqrt{g}}\right),
\label{secondary-special}
\end{align}
where
\begin{equation}
 \mathcal{H}_\perp \equiv \frac{2}{M_{\rm Pl}^2\sqrt{g}}\pi^{ij}\mathcal{G}_{ijkl}\pi^{kl}+\frac{M_{\rm Pl}^2}{2}\frac{\delta}{\delta N} \int \mathrm{d}^dx \sqrt{g}N \mathcal{L}_V,
\end{equation}
\begin{align}
 \phi_A \equiv 2\eta_0\sqrt{g} \left[ \frac{\lambda-1}{d\lambda-1}D^2\left(\frac{\pi}{\sqrt{g}}N\right) + R^{ij}\mathcal{G}_{ijkl}\frac{\pi^{kl}}{\sqrt{g}}N - D_iD_j\left(\frac{\pi^{ij}}{\sqrt{g}}N\right) \right],
  \label{eq:def-phi_A}
\end{align}
with $\pi \equiv g_{ij}\pi^{ij}$. They give secondary constraints
$J_A\approx 0$, $\mathcal{H}_i\approx 0$, $\mathcal{H}_\perp\approx 0$
and $\phi_A\approx 0$.

The set of constraints that we obtained so far is ($\pi_i$, $\pi_A$,
$\pi_\nu$, $\mathcal{H}_i$, $\pi_N$, $J_A$, $\mathcal{H}_\perp$,
$\phi_A$). Following the same logic as that in the generic case studied
in the previous subsection, we define the linear combination
$\mathcal{H}_i^N$ shown in Eq.~(\ref{eqn:def-HNi}). While the original 
constraint $\mathcal{H}_i$ has (weakly) nonvanishing Poisson brackets
with $\mathcal{H}_{\perp}$ and $\phi_A$ as 
\begin{align}
 \left\{\overline{\mathcal{H}}[f],\overline{\mathcal{H}_\perp}[\varphi]
 \right\}_{\rm P} 
 =& \overline{\mathcal{H}_\perp}[f\partial\varphi] +
 \int\mathrm{d}^dx
 \frac{\delta\overline{\mathcal{H}_\perp}[\varphi]}{\delta N}
 f^i\partial_iN
 \approx 
 \int\mathrm{d}^dx
 \frac{\delta\overline{\mathcal{H}_\perp}[\varphi]}{\delta N}
 f^i\partial_iN,
 \nonumber\\
 \left\{\overline{\mathcal{H}}[f],\overline{\phi_A}[\varphi]
 \right\}_{\rm P}
 =& \overline{\phi_A}[f\partial\varphi] + 
 \int\mathrm{d}^dx \frac{\delta\overline{\phi_A}[\varphi]}{\delta N}
 f^i\partial_iN
 \approx 
 \int\mathrm{d}^dx \frac{\delta\overline{\phi_A}[\varphi]}{\delta N}
 f^i\partial_iN,
 \label{eqn:HiHperp-HiphiA-special}
\end{align}
the Poisson brackets of $\mathcal{H}^N_i$ with all the constraints
vanish weakly: 
\begin{align}
 &\left\{\overline{\mathcal{H}^N}[f],\overline{\mathcal{H}^N}[g]\right\}_{\rm P} = \overline{\mathcal{H}^N}[[f,g]] \approx 0 \quad \text{for} \quad^\forall f^i, \ ^\forall g^i,
 \nonumber\\
 &\left\{\overline{\mathcal{H}^N}[f],\overline{\Phi}[\varphi]\right\}_{\rm P} = \overline{\Phi}[f\partial\varphi] \approx 0 \quad \text{for} \quad^\forall f^i, \ ^\forall \varphi,
 \label{eqn:HNi-HNi-Phi}
\end{align}
where $\Phi=\{\pi_N,J_A,\mathcal{H}_\perp,\phi_A\}$. Here, the formula
(\ref{eq:formula}) has been used to simplify the calculation of the
Poisson brackets shown in Eqs.~(\ref{eqn:HiHperp-HiphiA-special}) and
(\ref{eqn:HNi-HNi-Phi}). The Poisson brackets of $\pi_i$, $\pi_A$ and
$\pi_\nu$ with all the constraints vanish: 
\begin{equation}
 \left\{\pi_i(\x),\Phi(\vec{y})\right\}_{\rm P} = 0, \quad \left\{\pi_A(\x),\Phi(\vec{y})\right\}_{\rm P} = 0, \quad \left\{\pi_\nu(\x),\Phi(\vec{y})\right\}_{\rm P} = 0,
\end{equation}
where
$\Phi=\{\pi_j,\pi_A,\pi_\nu,\mathcal{H}^N_j,\pi_N,J_A,\mathcal{H}_\perp,\phi_A\}$. Finally,
while the Poisson brackets between $\pi_N$ and $J_A$ vanish,
\begin{align}
 &\left\{\pi_N(\x),\pi_N(\vec{y})\right\}_{\rm P} = \left\{\pi_N(\x),J_A(\vec{y})\right\}_{\rm P} = \left\{J_A(\x),J_A(\vec{y})\right\}_{\rm P} = 0,
  \label{eq:pi_N and J_A}
\end{align}
some of the Poisson brackets among $\pi_N$, $J_A$, $\mathcal{H}_\perp$
and $\phi_A$ are nonvanishing: 
\begin{align} \left\{\overline{\pi_N}[\chi],\overline{\mathcal{H}_\perp}[\varphi]\right\}_{\rm P} 
&= -\frac{M_{\rm Pl}^2}{2} \int\mathrm{d}^dx \chi \frac{\delta}{\delta N} \int\mathrm{d}^dy \varphi \frac{\delta}{\delta N} \int \mathrm{d}^dz \sqrt{g}N \mathcal{L}_V,
  \label{eq:pi_N and H_perp}\\
 \left\{\overline{\pi_N}[\chi],\overline{\phi_A}[\varphi]\right\}_{\rm P} 
 &= -2\eta_0\int\mathrm{d}^dx \chi \left(\frac{\lambda-1}{d\lambda-1}\pi D^2 + R^{ij}\mathcal{G}_{ijkl}\pi^{kl} - \pi^{ij}D_iD_j\right)\varphi,
  \label{eq:pi_N and phi_A}\\
 \left\{\overline{J_A}[\chi],\overline{\mathcal{H}_\perp}[\varphi]\right\}_{\rm P} 
 &\approx +2\eta_0\int\mathrm{d}^dx \varphi \left(\frac{\lambda-1}{d\lambda-1}\pi D^2 + R^{ij}\mathcal{G}_{ijkl}\pi^{kl} - \pi^{ij}D_iD_j\right)\chi,
  \label{eq:J_A and H_perp}\\
 \left\{\overline{J_A}[\chi],\overline{\phi_A}[\varphi]\right\}_{\rm P}
 &\approx 
 \eta_0^2M_{\rm Pl}^2\int\mathrm{d}^dx \sqrt{g}N
 \left[ R^{ij}\chi + (g^{ij}D^2-D^iD^j)\chi\right]
 \mathcal{G}_{ijkl}
 \left[ R^{kl}\varphi + (g^{kl}D^2-D^kD^l)\varphi\right].
 \label{eq:J_A and phi_A}
\end{align}

Hereafter we consider ($\pi_i$, $\pi_A$, $\pi_\nu$, $\mathcal{H}_i^N$,
$\pi_N$, $J_A$, $\mathcal{H}_\perp$, $\phi_A$) as the basis for the set
of constraints. Adding the terms corresponding to the secondary
constraints, we thus redefine the Hamiltonian as 
\begin{equation}
 \tilde{H} = H + \int \mathrm{d}^dx ( \lambda^i\pi_i+\lambda_A\pi_A+\lambda_\nu\pi_\nu+\lambda_\mathcal{H}^i\mathcal{H}_i^N
 +\lambda_N\pi_N+\lambda_JJ_A+\lambda_\perp\mathcal{H}_\perp+\lambda_\phi\phi_A),
\end{equation}
where $\lambda^i$, $\lambda_A$, $\lambda_\nu$, 
$\lambda_\mathcal{H}^i$, $\lambda_N$, $\lambda_J$, $\lambda_\perp$,
$\lambda_\phi$ are Lagrange multipliers.

The consistency conditions of $\pi_N$, $J_A$, $\mathcal{H}_\perp$ and $\phi_A$ with the time evolution yield no additional constraint but equations to determine the Lagrange multipliers $\lambda_\phi$, $\lambda_\perp$, $\lambda_J$ and $\lambda_N$ if the determinant of the following matrix does not vanish weakly \cite{Henneaux:1992}:
\begin{align}
 M&\equiv
  \left(
   \begin{array}{cccc}
    \left\{\overline{\pi_N}[\chi], \overline{\pi_N}[\varphi]\right\}_{\rm P} & \left\{\overline{\pi_N}[\chi], \overline{J_A}[\varphi]\right\}_{\rm P} & \left\{\overline{\pi_N}[\chi], \overline{\mathcal{H}_\perp}[\varphi]\right\}_{\rm P} & \left\{\overline{\pi_N}[\chi], \overline{\phi_A}[\varphi]\right\}_{\rm P}\\
    \left\{\overline{J_A}[\chi], \overline{\pi_N}[\varphi]\right\}_{\rm P} & \left\{\overline{J_A}[\chi], \overline{J_A}[\varphi]\right\}_{\rm P} & \left\{\overline{J_A}[\chi], \overline{\mathcal{H}_\perp}[\varphi]\right\}_{\rm P} & \left\{\overline{J_A}[\chi], \overline{\phi_A}[\varphi]\right\}_{\rm P}\\
    \left\{\overline{\mathcal{H}_\perp}[\chi], \overline{\pi_N}[\varphi]\right\}_{\rm P} & \left\{\overline{\mathcal{H}_\perp}[\chi], \overline{J_A}[\varphi]\right\}_{\rm P} & \left\{\overline{\mathcal{H}_\perp}[\chi], \overline{\mathcal{H}_\perp}[\varphi]\right\}_{\rm P} & \left\{\overline{\mathcal{H}_\perp}[\chi], \overline{\phi_A}[\varphi]\right\}_{\rm P}\\
    \left\{\overline{\phi_A}[\chi], \overline{\pi_N}[\varphi]\right\}_{\rm P} & \left\{\overline{\phi_A}[\chi], \overline{J_A}[\varphi]\right\}_{\rm P} & \left\{\overline{\phi_A}[\chi], \overline{\mathcal{H}_\perp}[\varphi]\right\}_{\rm P} & \left\{\overline{\phi_A}[\chi], \overline{\phi_A}[\varphi]\right\}_{\rm P}\\
   \end{array}
   \right).
\end{align}
With Eq.~\eqref{eq:pi_N and J_A}, $M$ can be partitioned with $2\times2$
matrices $B,C,D$ as 
\begin{align}
 M\equiv
   \begin{pmatrix}
     0 & B \\
     C & D \\
   \end{pmatrix},
\end{align}
where
\begin{equation}
 B=
   \begin{pmatrix}
     \left\{\overline{\pi_N}[\chi], \overline{\mathcal{H}_\perp}[\varphi]\right\}_{\rm P} & \left\{\overline{\pi_N}[\chi], \overline{\phi_A}[\varphi]\right\}_{\rm P} \\
     \left\{\overline{J_A}[\chi], \overline{\mathcal{H}_\perp}[\varphi]\right\}_{\rm P} & \left\{\overline{J_A}[\chi], \overline{\phi_A}[\varphi]\right\}_{\rm P}
      \end{pmatrix}, \quad
      C = -(B^T \mbox{ with } \chi \leftrightarrow \varphi),
\end{equation}
and $0$ is the $2\times 2$ matrix with all vanishing components. We thus have
\begin{equation}
 \det M = (\det B)(\det C) = 
  (\det B)(\det B \mbox{ with } \chi \leftrightarrow \varphi).
  \label{eq:det M}
\end{equation}
Equations \eqref{eq:pi_N and H_perp}--\eqref{eq:J_A and phi_A} show that each
component of the $2\times 2$ matrix $B$ is weakly
nonvanishing and that only Eq.~\eqref{eq:pi_N and H_perp} among them
depends on the potential term $\mathcal{L}_V[g_{ij},a_i]$, which is a
linear combination of $a^ia_i$, $R^{ij}a_ia_j$, $D^2a^iD^2a_i$ and so
forth. This means that the component \eqref{eq:pi_N and H_perp} linearly
depends on the coefficients of those potential terms but that other 
components are independent of them. Moreover, $\det B \not\approx0$ when
all coefficients in $\mathcal{L}_V[g_{ij},a_i]$ are set to
zero. Inspecting Eqs.~\eqref{eq:pi_N and H_perp}--\eqref{eq:J_A and phi_A}, we
thus conclude that $\det B \not\approx0$, for any choices of the
coupling constants in $\mathcal{L}_V$. Hence Eq.~\eqref{eq:det M} leads to 
$\det M\not\approx0$ and the consistency conditions of $\pi_N$, $J_A$,
$\mathcal{H}_\perp$ and $\phi_A$ do not yield any additional
constraints. 
The consistency conditions of $\pi_A$, $\pi_i$, $\pi_\nu$ and 
$\mathcal{H}^N_i$ with the time evolution are 
\begin{align}
 0 &= \frac{\mathrm{d}}{\mathrm{d}t}\pi_A \approx \left\{\pi_A,H\right\}_{\rm P} = -\frac{\delta H}{\delta A} = J_A,
 \nonumber\\
 0 &= \frac{\mathrm{d}}{\mathrm{d}t}\pi_i \approx \left\{\pi_i,H\right\}_{\rm P} = -\frac{\delta H}{\delta N^i} = -\mathcal{H}_i + J_AD_i\nu,
 \nonumber\\
 0 &= \frac{\mathrm{d}}{\mathrm{d}t}\pi_\nu \approx \left\{\pi_\nu,H\right\}_{\rm P} = -\frac{\delta H}{\delta \nu} = -\partial_i(N\mathcal{H}^i+\tilde{N}^iJ_A),
 \nonumber\\
 0 &= \frac{\mathrm{d}}{\mathrm{d}t}\overline{\mathcal{H}^N}[f] \approx \left\{\overline{\mathcal{H}^N}[f],H\right\}_{\rm P}
 = \int \dd^dx
 \left[
 -(N\mathcal{H}^i+\tilde{N}^iJ_A)\partial_i(f\partial\nu)-J_Af\partial A
 + (\mathcal{H}_i-J_A\partial_i\nu)[f,N]^i \right].
\end{align}
They do not give any additional constraints since the right-hand sides
are weakly vanishing.

Therefore, ($\pi_i$, $\pi_A$, $\pi_\nu$, $\mathcal{H}_i^N$, $\pi_N$,
$J_A$, $\mathcal{H}_\perp$, $\phi_A$) forms the set of all constraints
for the special case with $\eta_1=\eta_2=0$. Among them, ($2d+2$)
constraints ($\pi_i$, $\mathcal{H}_i^N$, $\pi_A$, $\pi_\nu$) are
first class and $4$ constraints ($\pi_N$, $J_A$, $\mathcal{H}_\perp$,
$\phi_A$) are second class.

\section{Number of degrees of freedom}
\label{sec:number of degrees of freedom}

\subsection{Generic case}
\label{subsec:dof-generic}

Our analysis in Sec.~\ref{sec:hamiltonian analysis} on the nonprojectable Ho\v{r}ava-Lifshitz theory with the $U(1)$ symmetry shows that we have $C_1 = 2d+2$ first-class constraints and $C_2=2$ second-class constraints in the generic case (Sec.~\ref{subsec:generic}) and $C_2 = 4$ in the special case (Sec.~\ref{subsec:special}) in the phase-space dimension ${\rm dim} P=d^2+3d+6$. Hence the number of degrees of freedom, $\mathcal{N}$, is
\begin{equation}
 \mathcal{N} = \frac{1}{2}(\mathrm{dim}P-2C_1-C_2) = \frac{1}{2}(d^2-d+2-C_2) \; .
\end{equation}
In the special case with $C_2=4$, ${\cal N} = \left(d-2\right)\left(d+1\right)/2$ counts the same as the number of transverse traceless polarizations of a massless graviton in $(d+1)$-dimensional spacetime, which shows the absence of the scalar graviton, the mode present in the original version of Ho\v{r}ava-Lifshitz gravity. On the other hand, in the generic nonprojectable case with $C_2=2$, there exists an additional degree of freedom in the theory, corresponding to a scalar graviton. One thus observes that despite an additional $U(1)$ symmetry, the scalar graviton generically remains present due to the lack of two second-class constraints, and the number of the propagating degrees of freedom is the same as that in the original Ho\v{r}ava-Lifshitz theory.

Since the first-class constraints $\pi_i \approx {\cal H}^N_i \approx \pi_A \approx \pi_\nu \approx 0$ are related to the generators of the gauge transformations under which the theory is invariant, the motion of the canonical variables is not completely determined by the Hamiltonian $\tilde H$ due to the unfixed multipliers $\lambda^i$, $\lambda_{\cal H}^i$, $\lambda_A$ and $\lambda_\nu$.
In order to remove this subtlety, one can in general fix the gauge by imposing additional constraints
\begin{equation}
    \mathcal{G}^i\approx 0, \ \mathcal{F}^i\approx 0,\ \mathcal{G}_A\approx 0,\ 
    \mathcal{G}_\nu \approx 0,\qquad (i=1,2,\dots,d),
\label{gauge-fixing}
\end{equation}
such that the determinant
\begin{equation}
 \det 
  \left(
   \begin{array}{cccc}
    \frac{\delta{\cal G}^i(\x)}{\delta N^j(\y)} & 
    \left\{{\cal G}^i(\x), {\cal H}^N_{j}(\y)\right\}_{\rm P} & 
    \frac{\delta{\cal G}^i(\x)}{\delta A(\y)} & \frac{\delta{\cal G}^i(\x)}{\delta \nu(\y)}\\
    \frac{\delta{\cal F}^i(\x)}{\delta N^j(\y)} &
     \left\{{\cal F}^i(\x), {\cal H}^N_{j}(\y)\right\}_{\rm P} &
    \frac{\delta{\cal F}^i(\x)}{\delta A(\y)} & \frac{\delta{\cal F}^i(\x)}{\delta \nu(\y)}\\
    \frac{\delta{\cal G}_A(\x)}{\delta N^j(\y)} & 
    \left\{{\cal G}_A(\x), {\cal H}^N_{j}(\y)\right\}_{\rm P} & 
    \frac{\delta{\cal G}_A(\x)}{\delta A(\y)} & \frac{\delta{\cal G}_A(\x)}{\delta \nu(\y)}\\
    \frac{\delta{\cal G}_\nu(\x)}{\delta N^j(\y)} & 
    \left\{{\cal G}_\nu(\x), {\cal H}^N_{j}(\y)\right\}_{\rm P} & 
    \frac{\delta{\cal G}_\nu(\x)}{\delta A(\y)} & \frac{\delta{\cal G}_\nu(\x)}{\delta \nu(\y)}\\
    \end{array}
   \right)
\end{equation} 
does not vanish weakly.
For the generic case $(\eta_1 ,\eta_2)\neq (0,0)$, we define the total Hamiltonian by including  these gauge-fixing conditions as
\begin{align}
\label{Htot}
H_{\rm tot} = \int {\rm d}^d x \Big[&
{\cal C} + N^i \left( {\cal H}_i - J_A D_i \nu \right) - A J_A
+ \lambda^i\pi_i + \lambda_A\pi_A+\lambda_\nu\pi_\nu 
+ \lambda_{\cal H}^i {\cal H}_i^N
\nonumber\\ &
+ \lambda_N\pi_N + \lambda_J J_A+ n_i{\cal G}^i + \lambda_i^{\cal F} {\cal F}^i + n_A{\cal G}_A +n_\nu{\cal G}_\nu
\Big]
\end{align}
where ($n_i,\,\lambda^\mathcal{F}_i,\, n_A,\, n_\nu$) are Lagrange multipliers, and we have defined
\begin{equation}
{\cal C} \equiv N \left[ \sqrt{g} \left( \frac{2}{M_{\rm Pl}^2} \, \frac{\pi^{ij}}{\sqrt{g}} {\cal G}_{ijkl} \frac{\pi^{kl}}{\sqrt{g}} 
+ \frac{M_{\rm Pl}^2}{2} {\cal L}_V \right)
- {\cal H}_i D^i \nu
+ \frac{1}{2} J_A D_i \nu D^i \nu \right] \; .
\label{def-C}
\end{equation} 
The set of all Lagrange multipliers ($\lambda^i$, $\lambda_A$, $\lambda_\nu$, $\lambda^i_{\mathcal{H}}$, $n_i$, $\lambda_{\mathcal{F}i}$, $n_A$, $n_\nu$, $\lambda_N$, $\lambda_J$) would be fully determined by imposing the consistency conditions on the constraints, which are now all second class, with $H_{\rm{tot}}$ instead of $\tilde{H}$.

As an explicit example of the gauge fixing, let us consider the following gauge:
\begin{align}
&\mathcal{G}^i = N^i, \quad\mathcal{F}^i = \mathcal{F}^i[g_{ij}, N, \nu, \pi^{kl}, \pi_N, \pi_\nu ], \nonumber \\
&\mathcal{G}_A = A, \;\quad \mathcal{G}_\nu = \mathcal{G}_\nu[g_{ij}, N, \nu, \pi^{kl}, \pi_N, \pi_\nu ],
\end{align}
for which the constraints $\mathcal{F}^i$ and $\mathcal{G}_\nu$ satisfy
\begin{equation}
\left\{{\cal F}^i(\x), {\cal H}^N_j(\y)\right\}_{\rm P} \not \approx 0,\quad 
\frac{\delta{\cal G}_\nu(\x)}{\delta \nu(\y)}\not \approx 0.
\label{cond-Fi-Gnu}
\end{equation}
In this case, the consistency conditions
\begin{equation}
\frac{d}{dt}\mathcal{G}^i \approx 0, \quad \frac{d}{dt}\mathcal{\pi}_i\approx 0,\quad 
\frac{d}{dt}\mathcal{G}_A\approx 0, \quad \frac{d}{dt}\mathcal{\pi}_A\approx 0 
\end{equation}
tell us that the $2d+2$ multipliers $(\lambda^i,\,\lambda_A,\,\, n_i,\, n_A)$ are determined as
\begin{equation}
\label{multipliers-GF}
   \lambda^i = 0,\quad  n_i = -\mathcal{H}_{i}+J_AD_i\nu, \quad \lambda_A =0,\quad n_A = J_A\ .
\end{equation}
Then the total Hamiltonian in this gauge is
\begin{equation}
H_{\rm tot} = \int {\rm d}^d x \left[ {\cal C} 
+\lambda_\nu\pi_\nu + \lambda_{{\cal H}}^i{\cal H}_i^N + \lambda_i^{\cal F} {\cal F}^i 
+ n_\nu{\cal G}_\nu +\lambda_N\pi_N
+ \lambda_J J_A\right] \; .
\end{equation}
From the absence of $(N^i,\,A,\,\pi_i,\,\pi_A)$ in this gauge-fixed total Hamiltonian, we can see that we have excluded the canonical pairs $(N^i,\,\pi_i)$ and $(A,\,\pi_A)$ from the phase space in this gauge. The dimension of the reduced phase space $(g_{ij},\, N,\,\nu,\,\pi^{ij},\, \pi_N, \pi_\nu)$ is $d^2+d+4$. As usual with second-class constraints, all the remaining Lagrange multipliers $(\lambda_\nu,\,\lambda_{\cal H}^i,\, \lambda_i^{\cal F},\, n_\nu,\, \lambda_N,\,\lambda_{J})$ are fully determined by imposing
\begin{alignat}{2}
 &\left\{\pi_\nu(\x), H_{\rm tot}\right\}_{\rm P} \approx 0, &\quad
&\left\{{\cal G}_\nu(\x), H_{\rm tot}\right\}_{\rm P} \approx 0,
 \nonumber\\
 &\left\{{\cal H}^{N}_{i}(\x), H_{\rm tot}\right\}_{\rm P} \approx 0, &\quad 
 &\left\{{\cal F}^i(\x), H_{\rm tot}\right\}_{\rm P} \approx 0,
 \nonumber\\
 &\left\{\pi_N(\x), H_{\rm tot}\right\}_{\rm P} \approx 0, &\quad 
&\left\{J_A(\x), H_{\rm tot}\right\}_{\rm P} \approx 0.
  \label{fixmultiplier-nonproject}
\end{alignat}
There remains the following set of $2d+4$ second-class constraints acting on the $(d^2+d+4)$-dimensional reduced phase space:
\begin{equation}
 \pi_\nu \approx 0, \quad {\cal G}_\nu \approx 0, \quad
 {\cal H}^N_{i} \approx 0, \quad   
 {\cal F}^i \approx 0, \quad 
 \pi_N \approx 0, \quad J_A \approx 0,  
 \qquad (i=1,2,\dots,d).
\end{equation}
Hence, in the generic case ($\eta_1 , \eta_2)\ne(0,0)$, we end up with a $(d^2-d)$-dimensional physical phase space, and the number of propagating degrees of freedom is $d(d-1)/2$, which shows that there exists a scalar graviton in the system. 

\subsection{Special case with $\eta_1=\eta_2=0$}
\label{subsec:dof-special}

The constraint structure and the number of degrees of freedom differ in a nontrivial way in the special case described in Sec.~\ref{subsec:special}, i.e. the case with ($\eta_1,\eta_2)=(0,0)$. Since $J_A$ is now independent of $N$, the time consistency conditions of $\pi_N$ and $J_A$ introduce two additional secondary constraints ${\cal H}_\perp$ and $\phi_A$, shown in Eq.~\eqref{secondary-special}, which are absent in the generic case. The total Hamiltonian contains these new constraint terms, and formally, we simply need to add two terms
\begin{equation}
\label{add}
\int {\rm d}^d x \left( \lambda_\bot {\cal H}_\bot + \lambda_\phi \phi_A \right)
\end{equation} 
to Eq.~(\ref{Htot}). We can fix the gauge and eliminate 
the canonical pairs $(N^i,\,\pi_i)$ and $(A,\,\pi_A)$ from the phase space by the same procedure as in the generic case.
In this special case, however, we have two additional second-class constraints,
${\cal H}_\bot \approx 0$ and $\phi_A \approx 0$. They reduce the dimension of the phase space by two, and the total number of propagating degrees of freedom becomes $(d^2-d-2)/2$, the same as the number of transverse traceless polarizations of a graviton. Hence this shows that the scalar graviton disappears in this special case, at least classically.

An apparent absence of the scalar graviton may arise in the perturbation theory even in the generic case. For example, if one sets $\eta_2 = 0$ in Eq.~\eqref{def-JA} and considers the vanishing background of $\sigma$ and the background of $N$ independent of the spatial coordinates, then the constraint structure appears to be the same as the special case up to the level of linear perturbations. This apparent behavior at the linear order is, however, only an artifact of the perturbative expansion around a specific background and does not mean that the system is free from a scalar graviton. In the situations of this sort, care must be taken, and one needs to consider the nonlinear orders of perturbations, at which the scalar graviton becomes dynamical.

\subsection{Projectable theory}
\label{subsec:projectable}

For completeness, we briefly discuss the projectable case, in which the lapse $N$ is a function only of time $t$. This restriction implies that the lapse is no longer a local degree of freedom, and thus we can consistently take it out of the phase space of the theory. As summarized in Appendix~\ref{sec:projectable}, the phase space consists of $\left( g_{ij}, N^i, A, \nu, \pi^{\ij}, \pi_i, \pi_A, \pi_\nu \right)$ with $2d+2$ first-class and $2$ second-class constraints, leading to $\left(d-2\right)\left(d+1\right)/2$ dynamical degrees of freedom. This equals the number of traceless transverse polarizations of a graviton, exhibiting the absence of the scalar graviton in the projectable case, in sharp contrast to the generic nonprojectable version of the theory (see Sec.~\ref{subsec:dof-generic}).

The total action, including the gauge-fixing conditions, of the projectable version of the $U(1)$ Ho\v{r}ava-Lifshitz theory is, from Eqs.~\eqref{def-H} and \eqref{Htil-projectable},
\begin{align}
H_{\rm tot}^{\rm proj} =& 
\int {\rm d}^d x \Big[{\cal C}^{\rm proj} + N^i \left( {\cal H}_i - J_A D_i \nu \right) - A J_A 
+ \lambda^i \pi_i + \lambda_A \pi_A + \lambda_\nu \pi_\nu 
\nonumber\\
& \qquad\quad\
+ \lambda_{\cal H}^i {\cal H}_i + \lambda_J J_A + \lambda_\phi \phi_A + n_i {\cal G}^i + \lambda^{\cal F}_i {\cal F}^i + n_A {\cal G}_A + n_\nu {\cal G}_\nu
\Big] \; ,
\end{align}
where ${\cal G}^i$,${\cal F}^i$, ${\cal G}_A$ and ${\cal G}_\nu$ are the gauge-fixing terms as in Eq.~\eqref{gauge-fixing}, and ${\cal C}^{\rm proj}$ is the projectable counterpart of ${\cal C}$, defined in Eq.~\eqref{def-C}. 
Note that the potential term ${\cal L}_V$ and the constraint $J_A$ now depend only on $g_{ij}$, since $\partial_i \ln N = 0$ and all of their dependence on $N$ vanishes in the projectable case.

Performing the gauge fixing in the same manner as in the previous subsection, we choose
\begin{eqnarray}
& {\cal G}^i = N^i \; ,&{\cal F}^i = {\cal F}^i \left[ g_{ij} , \nu , \pi^{kl} , \pi_\nu \right] \; ,
\nonumber\\
& {\cal G}_A = A \; , \; &{\cal G}_\nu = {\cal G}_\nu \left[ g_{ij} , \nu , \pi^{kl} , \pi_\nu \right] \; ,
\end{eqnarray}
imposing the conditions 
\begin{equation}
\left\{ {\cal F}^i (\x) , {\cal H}_j (\y) \right\} \not\approx 0 \; , \quad
\frac{\delta {\cal G}_\nu(\x)}{\delta\nu(\y)} \not\approx 0 \; ,
\label{cond-nonzero-proj}
\end{equation}
similar to Eq.~\eqref{cond-Fi-Gnu}. Then the time consistency conditions for ${\cal G}^i$, $\pi_i$, ${\cal G}_A$ and $\pi_A$ determine the multipliers $\left( \lambda^i , \lambda_A , n_i , n_A \right)$ in exactly the same way as in Eq.~\eqref{multipliers-GF}. After this gauge fixing, we can completely eliminate the gauge modes $N^i$, $A$, $\pi_i$ and $\pi_A$ from the phase space, resulting in the Hamiltonian
\begin{equation}
H_{\rm tot}^{\rm proj} =
\int {\rm d}^d x \Big[ {\cal C}^{\rm proj} + \lambda_\nu \pi_\nu 
+ \lambda_{\cal H}^i {\cal H}_i + \lambda_J J_A + \lambda_\phi \phi_A + \lambda^{\cal F}_i {\cal F}^i + n_\nu {\cal G}_\nu
\Big] \; .
\end{equation}
The evolution of the system with the reduced phase space $\left( g_{ij} , \nu , \pi^{ij} , \pi^\nu \right)$ is governed by this Hamiltonian. Due to the conditions in Eq.~\eqref{cond-nonzero-proj}, all of the constraints $\left( \pi_\nu , {\cal H}_i , J_A , \phi_A , {\cal F}^i , {\cal G}_\nu \right) \approx 0$ are now second class. Imposing them then reduces the phase-space dimension from $d(d+1)+2$ to $(d-2)(d+1)$, or equivalently $(d-2)(d+1)/2$ degrees of freedom, as desired. Similarly to Eq.~\eqref{fixmultiplier-nonproject}, the Lagrange multipliers are determined by the time consistency conditions for the constraints.

The Hamiltonian structure of the projectable case is closely similar to the special version of the nonprojectable case, described in Sec.~\ref{subsec:dof-special}. The only differences at the classical level are that the former excludes the lapse $N$ from the local degrees of freedom (and thus the corresponding second-class constraints are absent) and that ${\cal L}_V$ and $J_A$ in the former have no dependence on $N$. However, quantum effects are expected to detune such a special choice of parameters as in the special nonprojectable case, and fine-tunings would be required to keep those quantum corrections from arising. Thus even if one starts from the classical action without a scalar graviton in the nonprojectable theory, it would generically reappear at the quantum level. On the other hand, the projectable case is constructed in such a way that $N$ is constant at any constant-time hypersurface, and the absence of a scalar graviton is consistently guaranteed at the full order. The number of propagating degrees of freedom in the gravity sector is thus the same as that of a massless spin-$2$ graviton in GR.

\section{Discussion} 
\label{sec:discussion}

In this paper we have applied the standard method of Hamiltonian
analysis in the classical field theory to the $U(1)$ extension of
Ho\v{r}ava-Lifshitz gravity without the projectability condition. We
have studied the nature of constraints and counted the number of
physical degrees of freedom. In particular, we have shown that the
theory contains the scalar graviton unless the coefficients of
$a^ia_i\sigma$ and $D^ia_i\sigma$, which we denoted as $\eta_1$
and $\eta_2$, are set to zero exactly.

It is known that the scalar graviton is absent in linear perturbations
around a flat background even if $\eta_1\ne 0$, as far as
$\eta_2=0$. Actually, on a flat background the term $a^ia_i\sigma$ is of
the cubic order in perturbation and thus does not contribute to the
equations of motion for linear perturbations. This is the reason why the
linear perturbation analysis is insensitive to the value of $\eta_1$. On
the other hand, the absence of the scalar graviton at the fully nonlinear
level requires not only $\eta_2=0$ but also $\eta_1=0$.

The reason why $\eta_1$ and $\eta_2$ determine the presence/absence of
the scalar graviton is understood as follows. The Poisson bracket
between two constraints, that we denoted as $\pi_N$ and $J_A$, vanishes 
identically if and only if $\eta_1=\eta_2=0$. Thus, only in this special
case, the consistency of the two constraints with the time evolution
does not determine Lagrange multipliers but instead yields two secondary
constraints. These secondary constraints are responsible for the elimination
of the canonical pair corresponding to the scalar graviton.

We have identified the condition under which the scalar graviton is
absent, i.e $\eta_1=\eta_2=0$. However, the operators corresponding to
the two coupling constants are $a^ia_i\sigma$ and $D^ia_i\sigma$, and
both of them are marginal for any values of the dynamical critical 
exponent $z$. Therefore, even if $\eta_1$ and $\eta_2$ are set to zero
by hand, they should be generated by quantum corrections. In this sense
the condition $\eta_1=\eta_2=0$ is unstable under radiative
corrections. We thus conclude that the scalar graviton is generically
present in the theory.

Contrary to the generic nonprojectable theory, the projectable version
of the $U(1)$ extension of Ho\v{r}ava-Lifshitz gravity does not contain
the scalar graviton (see Sec.~\ref{subsec:projectable}). An
important point is that, unlike the above-mentioned condition
$\eta_1=\eta_2=0$ in the nonprojectable theory, the projectability
condition provides a consistent truncation and thus is expected to be
stable under radiative corrections. Hence, the number of physical
degrees of freedom in the projectable theory is the same as in GR.\footnote{Lovelock's theorem~\cite{Lovelock:1971yv} is
evaded because of the presence of auxiliary fields and nontrivial
constraints.} Nonetheless, physical properties of the propagating
degrees of freedom, e.g. the dispersion relation, are different from
those in GR. It would certainly be of theoretical
interest to determine whether there are any other such theories.

\section*{Acknowledgements}

The authors thank C.~M.~Melby-Thompson and T.~Sotiriou for useful
discussions. S. M. was supported in part by Grant-in-Aid for Scientific
Research 24540256, and Y. W. was supported by the Program for Leading
Graduate Schools, MEXT, Japan. The authors acknowledge support by the
WPI Initiative, MEXT Japan. Part of this work has been done within the
Labex ILP (reference ANR-10-LABX-63) part of the Idex SUPER, and
received financial state aid managed by the Agence Nationale de la
Recherche, as part of the programme Investissements d'avenir under the
reference ANR-11-IDEX-0004-02. The authors are thankful to Institut
Astrophysique de Paris for warm hospitality during their stay.

\appendix
\appendixpage

\section{Calculation of $\left\{\overline{\mathcal{H}}[f],F\right\}_{\rm P}$}
\label{sec:useful relations}

In this appendix we calculate the Poisson bracket
$\left\{\overline{\mathcal{H}}[f],F\right\}_{\rm P}$, where
$F=F[g_{ij},\pi^{ij},s,V^i]$ is a functional of the spatial metric
$g_{ij}$, its canonical momenta $\pi^{ij}$, a scalar $s$ and a vector
$V^i$ on a constant-$t$ surface and we assume that $F$ is invariant
under time-independent spatial diffeomorphism. We do {\it not} require
the invariance of $F$ under a time-dependent spatial diffeomorphism 
{\it nor} assume any properties of $s$ and $V^i$ under the time-dependent
spatial diffeomorphism.

We consider the time-independent spatial diffeomorphism,
\begin{equation}
 \vec{x} \to \vec{x} + \vec{\xi}(\vec{x}),
\end{equation}
where $\vec{\xi}(\vec{x})$ is a time-independent $d$-dimensional spatial
vector. Under this infinitesimal transformation, the spatial metric
$g_{ij}$, its canonical momenta $\pi^{ij}$, a scalar $s$ and a vector
$V^i$ transform as
\begin{eqnarray}
 g_{ij} & \to & g_{ij} + D_i\xi_j+D_j\xi_i, \nonumber\\
 \frac{\pi^{ij}}{\sqrt{g}} & \to & \frac{\pi^{ij}}{\sqrt{g}}
  + \xi^kD_k\left(\frac{\pi^{ij}}{\sqrt{g}}\right)
  - \frac{\pi^{ik}}{\sqrt{g}}D_k\xi^j
  - \frac{\pi^{kj}}{\sqrt{g}}D_k\xi^i,  \nonumber\\
 s & \to & s + \xi^k\partial_k s, \nonumber\\
 V^i & \to & V^i + \xi^kD_kV^i - V^kD_k\xi^i,
\end{eqnarray}
where $D_k$ is the spatial covariant derivative compatible with
$g_{ij}$ and we have used the fact that $\pi^{ij}/\sqrt{g}$ is a
tensor. The variation of a functional $F$ of these variables is thus
calculated as
\begin{align}
 \delta F =& 
 \int \mathrm{d}^dx 
 \left[ \left(\frac{\delta F}{\delta g_{ij}}\right)_{\pi/\sqrt{g}}
 \delta g_{ij} + \frac{\delta F}{\delta (\pi^{ij}/\sqrt{g})} \delta
 \left( \frac{\pi^{ij}}{\sqrt{g}} \right) + \frac{\delta F}{\delta s}
 \delta s + \frac{\delta F}{\delta V^i} \delta V^i \right] 
 \nonumber\\
 =& 
 \int \mathrm{d}^dx \xi^i 
 \biggl\{ -2g_{ik}\sqrt{g}D_j
 \left[\frac{1}{\sqrt{g}}
 \left(\frac{\delta F}{\delta g_{kj}}\right)_{\pi/\sqrt{g}}\right] 
 + \frac{\delta F}{\delta (\pi^{jk}/\sqrt{g})}D_i
 \left(\frac{\pi^{jk}}{\sqrt{g}}\right)
 \nonumber\\
 & + 2\sqrt{g}D_j\left(\frac{1}{\sqrt{g}}\frac{\delta F}{\delta(\pi^{il}/\sqrt{g})}\frac{\pi^{jl}}{\sqrt{g}}\right) + \frac{\delta F}{\delta s}D_is  + \frac{\delta F}{\delta V^j}D_iV^j + \sqrt{g}D_j\left(\frac{1}{\sqrt{g}}\frac{\delta F}{\delta V^i}V^j\right)
 \biggr\},
\end{align}
where the subscript $\pi/\sqrt{g}$ in 
$(\delta F/\delta g_{ij})_{\pi/\sqrt{g}}$ indicates that the functional
derivative is taken with $\pi^{kl}/\sqrt{g}$ (instead of $\pi^{kl}$)
fixed. Therefore, the diffeomorphism invariance of $F$, i.e. 
$\delta F=0$ for ${}^{\forall}\xi^i$, implies that 
\begin{align}
 2D_j 
 \left[\frac{1}{\sqrt{g}}
 \left(\frac{\delta F}{\delta g_{ij}}\right)_{\pi/\sqrt{g}}\right] 
 = & 
 \frac{1}{\sqrt{g}}\frac{\delta F}{\delta (\pi^{jk}/\sqrt{g})}D^i
 \left(\frac{\pi^{jk}}{\sqrt{g}}\right)
+ 2g^{ik}D_j\left(\frac{1}{\sqrt{g}}
 \frac{\delta F}{\delta(\pi^{kl}/\sqrt{g})}
 \frac{\pi^{jl}}{\sqrt{g}}\right)
 \nonumber\\
 & 
 + \frac{1}{\sqrt{g}}\frac{\delta F}{\delta s}D^is 
 + \frac{1}{\sqrt{g}}\frac{\delta F}{\delta V^j}D^iV^j
 + g^{ik}D_j
 \left(\frac{1}{\sqrt{g}}\frac{\delta F}{\delta V^k}V^j\right).
\label{eq:useful relation1}
\end{align}

For practical purposes, it is convenient to express 
$(\delta F/\delta g_{ij})_{\pi/\sqrt{g}}$ in terms of 
$(\delta F/\delta g_{ij})_\pi$, where 
the subscript $\pi$ indicates that the functional derivative is taken
with $\pi^{kl}$ (instead of $\pi^{kl}/\sqrt{g}$) fixed. 
By writing down the variation $\delta F$ in two different ways as
\begin{equation}
\left(\frac{\delta F}{\delta g_{ij}}\right)_\pi \delta g_{ij}
 + \frac{\delta F}{\delta \pi^{ij}}\delta \pi^{ij} 
 + \frac{\delta F}{\delta s}\delta s 
  + \frac{\delta F}{\delta V^i}\delta V^i 
 = \left(\frac{\delta F}{\delta g_{ij}}\right)_{\pi/\sqrt{g}} \delta g_{ij}
 + \frac{\delta F}{\delta (\pi^{ij}/\sqrt{g})}
 \delta\left(\frac{\pi^{ij}}{\sqrt{g}}\right)
 + \frac{\delta F}{\delta s}\delta s 
  + \frac{\delta F}{\delta V^i}\delta V^i,
\end{equation}
and equating the coefficients of $\delta g_{ij}$ and $\delta\pi^{ij}$,
we obtain
\begin{equation}
 \left(\frac{\delta F}{\delta g_{ij}}\right)_{\pi/\sqrt{g}} 
 = 
 \left(\frac{\delta F}{\delta g_{ij}}\right)_\pi 
 + \frac{1}{2}\frac{\delta F}{\delta\pi^{kl}}\pi^{kl}g^{ij},
 \quad
 \frac{\delta F}{\delta (\pi^{ij}/\sqrt{g})}
 = 
 \sqrt{g}\frac{\delta F}{\delta \pi^{ij}}.
\label{eq:useful relation2}
\end{equation}
By substituting Eq.~\eqref{eq:useful relation2} into Eq.
\eqref{eq:useful relation1}, we thus obtain
\begin{align}
 2D_j 
 \left[\frac{1}{\sqrt{g}}
 \left(\frac{\delta F}{\delta g_{ij}}\right)_\pi\right] 
 = & 
 - D^i
 \left( \frac{\delta F}{\delta \pi^{kl}}\frac{\pi^{kl}}{\sqrt{g}}
 \right)
 + \frac{\delta F}{\delta \pi^{jk}}D^i
 \left(\frac{\pi^{jk}}{\sqrt{g}}\right)
+ 2g^{ik}D_j
 \left( \frac{\delta F}{\delta\pi^{kl}}\frac{\pi^{jl}}{\sqrt{g}}\right)
 \nonumber\\
 & 
 + \frac{1}{\sqrt{g}}\frac{\delta F}{\delta s}D^is 
 + \frac{1}{\sqrt{g}}\frac{\delta F}{\delta V^j}D^iV^j
 + g^{ik}D_j
 \left(\frac{1}{\sqrt{g}}\frac{\delta F}{\delta V^k}V^j\right).
 \label{eq:useful relation}
\end{align}
This formula will be used below.

For the calculation of 
$\left\{\overline{\mathcal{H}}[f],F\right\}_{\rm P}$, we first need to
calculate the functional derivatives of $\mathcal{H}[f]$. A
straightforward calculation leads to
\begin{align}
 \frac{1}{\sqrt{g}}\frac{\delta\bar{\mathcal{H}}[f]}{\delta g_{ij}}
 & = 
 \frac{\pi^{ik}}{\sqrt{g}}D_kf^j
 + \frac{\pi^{kj}}{\sqrt{g}}D_kf^i
 - D_k\left(\frac{\pi^{ij}}{\sqrt{g}}f^k\right), \nonumber\\
 \frac{\delta\bar{\mathcal{H}}[f]}{\delta \pi^{ij}} 
 & = 
 D_if_j + D_jf_i.
\end{align}
Hence, we obtain
\begin{align}
 \left\{\overline{\mathcal{H}}[f],F\right\}_{\rm P} 
 &= 
 \int \mathrm{d}^dx 
 \left[
 \frac{\delta\bar{\mathcal{H}}[f]}{\delta g_{ij}}
 \frac{\delta F}{\delta \pi^{ij}} 
 - 
 \frac{\delta\bar{\mathcal{H}}[f]}{\delta \pi^{ij}} 
 \left(\frac{\delta F}{\delta g_{ij}}\right)_\pi
 \right] \nonumber\\
 & =
 \int \mathrm{d}^dx \sqrt{g}
 \left\{
 \left[
 \frac{\pi^{ik}}{\sqrt{g}}D_kf^j
 + \frac{\pi^{kj}}{\sqrt{g}}D_kf^i
 - D_k\left(\frac{\pi^{ij}}{\sqrt{g}}f^k\right)
 \right]
 \frac{\delta F}{\delta \pi^{ij}} 
 + 2f_iD_j
 \left[
 \frac{1}{\sqrt{g}}
 \left(\frac{\delta F}{\delta g_{ij}}\right)_\pi
 \right]
 \right\}.
\end{align}
By using Eq.~(\ref{eq:useful relation}), we thus obtain
\begin{equation}
 \left\{\overline{\mathcal{H}}[f],F\right\}_{\rm P} 
 =\int \mathrm{d}^dx 
 \left( \frac{\delta F}{\delta s}f^i\partial_is 
  + \frac{\delta F}{\delta V^i}[f,V]^i \right).
\end{equation}
Here, $f^i$ is assumed to be independent of canonical variables. If
$f^i$ depends on canonical variables then we instead obtain 
\begin{equation}
 \left\{\overline{\mathcal{H}}[f],F\right\}_{\rm P} 
 =
 \overline{\mathcal{H}}[ \left\{f,F\right\}_{\rm P}]
 +
 \int \mathrm{d}^dx 
 \left( \frac{\delta F}{\delta s}f^i\partial_is 
  + \frac{\delta F}{\delta V^i}[f,V]^i \right)
 \approx
 \int \mathrm{d}^dx 
 \left( \frac{\delta F}{\delta s}f^i\partial_is 
  + \frac{\delta F}{\delta V^i}[f,V]^i \right).
\end{equation}

\section{Hamiltonian analysis of the projectable theory}
\label{sec:projectable}

In order to compare the analysis in the present paper with that under
the projectable condition $N=N(t)$, i.e.\ $a_i=0$, performed in Ref. \cite{Kluson:2010zn}, in this appendix we sketch the analysis for the
projectable theory using our notation. In the projectable case, the
lapse $N$ is not a local degree of freedom and the Hamiltonian
constraint, i.e. the equation of motion for $N(t)$, is not a local
equation but an equation integrated over the whole space. Since the whole
space here includes not only our patch of the universe but also many
other patches, the Hamiltonian constraint of this form does not restrict
the behavior of local degrees of freedom~\cite{Mukohyama:2009mz}. We
thus do not have to impose the Hamiltonian constraint to count the
number of  degrees of freedom. In the following, we thus simply consider
$N(t)$ as a fixed positive function of $t$ and we do not vary it. The
phase space ($g_{ij}$, $\pi^{ij}$, $N^i$, $\pi_i$, $A$, $\pi_A$, $\nu$,
$\pi_\nu$) is $(d^2+3d+4)$ dimensional.

The primary constraints are
\begin{equation}
 \pi_i=0, \quad \pi_A=0, \quad \pi_\nu+J_A=0,
\end{equation}
whose Poisson brackets all vanish. The consistency conditions of the primary constraints are
\begin{align}
 0 &= \frac{\mathrm{d}}{\mathrm{d}t}\pi_A \approx \left\{\pi_A,H\right\}_{\rm P} = -\frac{\delta H}{\delta A} = J_A,
 \nonumber\\
 0 &= \frac{\mathrm{d}}{\mathrm{d}t}\pi_i \approx \left\{\pi_i,H\right\}_{\rm P} = -\frac{\delta H}{\delta N^i} = -\mathcal{H}_i+ J_AD_i\nu,
 \nonumber\\
 0 &= \frac{\mathrm{d}}{\mathrm{d}t}(\pi_\nu+J_A) \approx - \frac{\delta H}{\delta \nu} + \left\{J_A,H\right\}_{\rm P}
 \nonumber\\
 &= -\partial_i(N\mathcal{H}^i+\tilde{N}^iJ_A) + \phi_A + \frac{2}{M_{\rm Pl}^2\sqrt{g}}g^{ij}\mathcal{G}_{ijkl}\pi^{kl}NJ_A + \sqrt{g}D_i(\tilde{N}^iJ_A/\sqrt{g}).
\end{align}
So far the constraints are ($\pi_i$, $\pi_A$, $\pi_\nu$, $\mathcal{H}_i$, $J_A$, $\phi_A$). Only the Poisson bracket between $J_A$ and $\phi_A$ does not vanish weakly [Eq.~\eqref{eq:J_A and phi_A}] among them. We redefine the Hamiltonian as
\begin{align}
 \tilde{H} = H + \int \mathrm{d}^dx (\lambda^i\pi_i + \lambda_A\pi_A + \lambda_\nu\pi_\nu + \lambda_\mathcal{H}^i\mathcal{H}_i + \lambda_JJ_A + \lambda_\phi\phi_A),
  \label{Htil-projectable}
\end{align}
where $\lambda^i$, $\lambda_A$, $\lambda_\nu$, $\lambda_\mathcal{H}^i$, $\lambda_J$, $\lambda_\phi$ are Lagrange multipliers.

The consistency conditions of $J_A$ and $\phi_A$ with the time evolution
do not yield additional constraints but simply result in equations to
determine the Lagrange multipliers $\lambda_J$ and $\lambda_\phi$. The
consistency conditions of $\pi_i$, $\pi_A$, $\pi_\nu$ and
$\mathcal{H}_i$ with the time evolution are 
\begin{align}
 0 &= \frac{\mathrm{d}}{\mathrm{d}t}\pi_A \approx \left\{\pi_A,H\right\}_{\rm P} = -\frac{\delta H}{\delta A} = J_A,
 \nonumber\\
 0 &= \frac{\mathrm{d}}{\mathrm{d}t}\pi_i \approx \left\{\pi_i,H\right\}_{\rm P} = -\frac{\delta H}{\delta N^i} = -\mathcal{H}_i+ J_AD_i\nu,
 \nonumber\\
 0 &= \frac{\mathrm{d}}{\mathrm{d}t}\pi_\nu \approx \left\{\pi_\nu,H\right\}_{\rm P} = -\frac{\delta H}{\delta \nu} = -\partial_i(N\mathcal{H}^i+\tilde{N}^iJ_A),
 \nonumber\\
 0 &= \frac{\mathrm{d}}{\mathrm{d}t}\overline{\mathcal{H}}[f] \approx \left\{\overline{\mathcal{H}}[f],H\right\}_{\rm P}
 = \int \dd^dx
 \left[
 -(N\mathcal{H}^i+\tilde{N}^iJ_A)\partial_i(f\partial\nu)-J_Af\partial A
 + (\mathcal{H}_i-J_A\partial_i\nu)[f,N]^i \right].
\end{align}
They weakly vanish and thus do not give any additional
constraints. Therefore ($\pi_i$, $\mathcal{H}_i$, $\pi_A$, $\pi_\nu$)
are ($2d+2$) first-class constraints and $J_A$ and $\phi_A$ are two
second-class constraints. The number degrees of freedom $\mathcal{N}$ is 
\begin{equation}
 \mathcal{N} = \frac{1}{2}\left[(d^2+3d+4)-2(2d+2)-2\right] = \frac{1}{2}(d^2-d-2),
\end{equation}
which implies the absence of the scalar graviton.

\bibliographystyle{apsrmp}
\bibliography{rmp-sample}

\end{document}